\documentclass[final,5p,times,twocolumn]{elsarticle}

 \usepackage{graphics,epsfig,subfigure}

\usepackage{amsmath}
\usepackage{amssymb}
\usepackage{cases}
\usepackage{hyperref}
\usepackage{xcolor}

\newcommand{\be}{\begin{equation}}
\newcommand{\ee}{\end{equation}}
\newcommand{\beq}{\begin{eqnarray}}
\newcommand{\eeq}{\end{eqnarray}}
\allowdisplaybreaks
\biboptions{comma,square,compress,numbers,sort}

\journal{Physics of Dark Universe}

\begin{document}

\begin{frontmatter}



\title{Inflation with shallow dip and primordial black holes}


\author[label1,label2]{Bao-Min Gu}
  \ead{gubm@ncu.edu.cn}
\author[label1,label2,label3]{Fu-Wen Shu}
   \ead{shufuwen@ncu.edu.cn}
\author[label4]{Ke Yang\corref{cor1}}
  \ead{keyang@swu.edu.cn}
  \cortext[cor1]{The corresponding author.}

\address[label1]{Center for Relativistic Astrophysics and High Energy Physics, Nanchang University, Nanchang, 330031, China}
\address[label2]{Center for Relativistic Astrophysics and High Energy Physics, Nanchang University, Nanchang 330031, China}
\address[label3]{Center for Gravitation and Cosmology, Yangzhou University, Yangzhou, 225009, China}
\address[label4]{School of Physical Science and Technology, Southwest University, Chongqing 400715, China}

\begin{abstract}

Primordial black holes may arise through ultra slow-roll inflation. In this work we study a toy model of ultra slow-roll inflation with a shallow dip. The ultra slow-roll stage enhances the curvature perturbations and thus the primordial scalar power spectrum. We analyze the features of the power spectrum numerically and analytically, and then give a rough estimate of the lower and upper bound of the enhancement. These large perturbations also produce second order gravitational waves, which are in the scope of future observations.

\end{abstract}


\end{frontmatter}



\section{Introduction}
Inflationary cosmology has great success in solving the puzzles of standard big bang theory. In addition, it provides elegant explanations for the temperature fluctuations of the cosmic microwave background (CMB) and the seeds for large scale structure (LSS) formation. Present observations of CMB \cite{Planck:2018jri} and LSS \cite{SDSS:2003eyi,SDSS-III:2015hof,DES:2017myr,Planck:2018vyg} have significantly constrained inflation on scales larger than several Gpc.
However, the constraints on small scales are much more weaker, leaving the details of inflationary dynamics on these scales unknown.

A bold but intriguing postulation is that primordial black holes (PBHs) may arise in very early universe \cite{Zeldovich1966,Hawking:1971ei,Carr:1974nx}.
These black holes are a type of hypothetical objects formed from the collapse of overdense region caused by large perturbations on small scales. Due to the primordial origin, they have a wide range of mass down to the Planck mass ($10^{-5}$ g) and up to as heavy as supermassive black holes. The PBHs lighter than $\sim10^{15}$g have evaporated through Hawking radiation \cite{Hawking:1974rv,Hawking:1975vcx}, while those heavier still survive at present epoch. These relic PBHs are natural dark matter candidate because they are collision-less and friction-less. They are also considered as the sources of the gravitational wave signals observed by LIGO/Virgo/KAGRA \cite{Sasaki:2016jop,Raidal:2017mfl,Ali-Haimoud:2017rtz,Raidal:2018bbj,Vaskonen:2019jpv,
DeLuca:2020bjf,DeLuca:2020qqa,Clesse:2020ghq,Hall:2020daa,DeLuca:2020sae,Wong:2020yig,
Hutsi:2020sol,DeLuca:2021wjr,Franciolini:2021tla,Franciolini:2021xbq,Franciolini:2022iaa}. Current data from various channels of observations have imposed severe constraints on the mass (energy) fraction of PBHs \cite{Ivanov:1994pa,Carr:2009jm,Saito:2009jt,Barnacka:2012bm,Capela:2013yf,Carr:2016drx,
Niikura:2017zjd,Carr:2017jsz,Zumalacarregui:2017qqd,Sasaki:2018dmp,Niikura:2019kqi,
Montero-Camacho:2019jte,Laha:2019ssq,Dasgupta:2019cae,
Carr:2020gox,Carr:2020xqk,Green:2020jor,Mittal:2021egv,Ozsoy:2023ryl}. For example, the mass range $10^{5}\sim10^{12}~M_{\odot}$ is strongly limited by the spectral distortion of the CMB \cite{Fixsen:1996nj,Chluba:2012we,Kohri:2014lza,Nakama:2017xvq}, the range $10^{-3}\sim1~M_{\odot}$ by Pulsar Timing Arrays \cite{Chen:2019xse}, where $M_{\odot}=2\times 10^{33}g$ is the solar mass. These are indirect constraints that rely on some model details. Some direct constraints like gravitational lensing \cite{Niikura:2017zjd,Diego:2017drh,Oguri:2017ock}, constrains the mass range $10^{-13}\sim10^{-5}~M_{\odot}$ stringently. In most mass windows, the fraction of PBHs as dark matter is less than $1\%$. However, there is still an asteroid mass window, $10^{-16}\sim10^{-13}~M_{\odot}$, in which the PBHs could contribute a large fraction or the total of dark matter.

The most commonly considered mechanism of generating the large perturbations required for PBH formation is inflation, which provides the origin of the primordial quantum fluctuations naturally. To have the PBHs formed, the curvature perturbations need to be enhanced sufficiently. For gaussian distributed curvature perturbations, the primordial scalar power spectrum at scales related to the PBH formation is required to be roughly seven orders larger than that at CMB scales \cite{Young:2014ana}. This may change if the details like non-Gaussianities and non-linear effects are considered \cite{Young:2013oia,Pattison:2017mbe,Franciolini:2018vbk,Biagetti:2018pjj,Ezquiaga:2018gbw,
Atal:2018neu,Passaglia:2018ixg,Young:2019yug,Young:2019osy,Kehagias:2019eil,Ezquiaga:2019ftu,
Kitajima:2021fpq,Figueroa:2021zah,Cai:2021zsp,Cai:2022erk,Young:2022phe,Ferrante:2022mui,
Pi:2022ysn,vanLaak:2023ppj}. The enhancement can be realised in various inflation models, of which one of the  simplest is the single field inflation with an ultra slow-roll (USR) stage. Usually, the USR phase appears if the potential has, for instance, a quasi-inflection point \cite{Ivanov:1994pa,Yokoyama:1998pt,Cheng:2016qzb,Garcia-Bellido:2017mdw,Ezquiaga:2017fvi,Germani:2017bcs,
Germani:2017bcs,Cicoli:2018asa,Ozsoy:2018flq,Cicoli:2022sih}, a local minimum/maximum \cite{Kannike:2017bxn,Ballesteros:2017fsr,Hertzberg:2017dkh,Biagetti:2018pjj,Mishra:2019pzq,
Figueroa:2020jkf,
Karam:2022nym,Gu:2022pbo}, etc. When the inflaton rolls over these features, the slow roll phase transitions to be USR and the inflaton is largely decelerated. In such a stage the curvature perturbations and the primordial power spectrum would be amplified. The PBHs may form after these amplified perturbations re-enter the horizon, provided that the enhancement is sufficient and the USR phase is long enough. The enhancement mechanism is also widely studied in inflation models constructed with piecewise potentials \cite{Kefala:2020xsx,Inomata:2021uqj,Dalianis:2021iig,Inomata:2021tpx}. This type of theories are useful to show the details of the amplification mechanism. However, the non-smooth features may lead to sharp and instantaneous transitions. As is discussed recently \cite{Kristiano:2022maq,Riotto:2023hoz,Choudhury:2023vuj,Choudhury:2023jlt,Kristiano:2023scm,
Choudhury:2023rks,Firouzjahi:2023ahg,Franciolini:2023lgy,Fumagalli:2023hpa}, such transitions may have significant loop corrections from the small scales to the power spectrum at large scales and spoil the CMB observations.
\begin{table*}
\begin{center}
\begin{tabular}{c|c |c| c |c |c| c |c}
\hline\hline
Sets &$\phi_{\text{CMB}}[M_p]$ &$\alpha [M_p^4]$& $\xi [M_p^4]$ &$\upsilon_1[M_p]$& $\upsilon_2[M_p]$ & $\upsilon_3[M_p]$
 \\ [0.5ex] %
\hline
$P_1$  &0.1912& $1.425\times 10^{-15}$ & $1.425\times 10^{-13}$ &  $0.094$ & $0.02446015$ & $0.0099$    \\
$P_2$ &0.1930& $1.933\times 10^{-15}$ & $1.740\times 10^{-13}$ &  0.096 & 0.0254076 & 0.01\\
$P_3$ &0.1920& $1.530\times 10^{-15}$ & $1.530\times 10^{-15}$ &  0.095 & 0.024643 & 0.01 \\
$P_4$ &0.1935& $1.530\times 10^{-15}$ & $1.530\times 10^{-15}$ &  0.098 & 0.02569 & 0.01
\\
 [0.5ex] 
\hline\hline
\end{tabular}
\end{center}
\caption{Parameter sets used in this paper. We set $\lambda=5\times 10^6 M_p^{-4}$ for all the sets. Since the PBH abundance is sensitive to the power spectrum, the parameter $\upsilon_2$ is fine-tuned so that the PBH abundance is of interest and not overproduced.}
\label{table1}
\end{table*}
Hence it is of great importance to consider smooth and more realistic models.
One may also consider non-standard theories like non-canonical scalar field \cite{Cai:2018tuh,Ballesteros:2018wlw,Kamenshchik:2018sig,Lin:2020goi,Chen:2020uhe,
Yi:2020cut,Solbi:2021wbo,Teimoori:2021pte}, modified gravity \cite{Pi:2017gih,Fu:2019ttf,Heydari:2021gea,Kawai:2021edk,Frolovsky:2022ewg}, multi-field inflation \cite{Garcia-Bellido:1996mdl,Palma:2020ejf,Braglia:2020eai,Cai:2021wzd,
Hooshangi:2022lao,Geller:2022nkr}, etc.

In this work we consider the non-monotonic model proposed in \cite{Gu:2022pbo}. The potential has a shallow dip, with a local minimum and a local maximum. We study the primordial power spectrum both numerically and analytically. In particular, we try to improve the analytical expression of the power spectrum. The analytical description of the power spectrum \cite{Chongchitnan:2006wx,Motohashi:2017kbs,Byrnes:2018txb,Liu:2020oqe,Tasinato:2020vdk,Karam:2022nym} is meaningful since it relates the model parameters or the slow-roll parameters to the PBH abundance directly. If the analytical expression is highly precise, it would be instructive for building models. We also study the second order gravitational waves (GWs) produced by the large perturbations \cite{Ananda:2006af,Baumann2007,Saito:2008jc,Bugaev:2009zh,Kohri:2018awv,Cai:2018dig,
Inomata:2018epa,Domenech:2021ztg}. These GWs are in the scope of future observatories. The detection of stochastic GWs background would imply new physics of the universe. However, the non-detection would also impose strong constraints on the early universe. Compared to \cite{Gu:2022pbo}, We used the same model and the mechanism of USR, but with different parameter sets, which are related to the asteroid mass PBHs. The improvements in this paper are primarily the analysis of the scalar power spectrum and the gravitational waves induced by scalar perturbations. Although our results do not provide an explicit relationship between the power spectrum and the model parameters, we believe our study enhances the understanding of how ultra slow-roll inflation amplifies curvature perturbations, leading to PBH formation.

This paper is organized as follows. We first introduce the model briefly in section \ref{sec1}, then we solve the perturbations and give the numerical and analytical power spectrum in section \ref{sec2}. The PBH formation and the induced gravitational waves are considered in section \ref{sec3} and \ref{sec4}, respectively. The summary is given in the final section.

\section{The model}\label{sec1}

The model proposed in \cite{Gu:2022pbo} contains two parts,
\be
V(\phi)=\alpha \frac{f_1(\phi)}{1+f_1(\phi)}+\xi \frac{f_2(\phi)}{1+f_2(\phi)},
\label{inflationpotential}
\ee
where $\phi$ is the inflaton field, $f_1=\frac{1}{4}\lambda \phi^2\left((\phi-\upsilon_1)^2+\upsilon_2^2\right)$, and $f_2=\phi^4/\upsilon_3^4$. There are six parameters in our model, $\alpha$, $\xi$, $\lambda$, $v_1$, $v_2$, and $v_3$. The first part of the potential in Equation (\ref{inflationpotential}) provides the feature, which is controlled by the parameter $\alpha$. The second part is a standard slow-roll potential proportional to the parameter $\xi$, which is constrained by the CMB observations by using $V(\phi_{CMB})\simeq\alpha+\xi$ (with $\alpha<<\xi$). The parameter $\alpha$ controls the size of the feature, and cannot be severely constrained. To have a successful ultra slow-roll inflation model, $\alpha$ must be sufficiently small. In our model we set $\alpha\sim0.01\xi$. Since $\alpha<<\xi$, the parameter $v_3$ can also be constrained by using the constraints on $n_s$. The remaining parameters $\lambda$, $v_1$, and $v_2$ are relatively free. These free parameters describe the characteristics of the dip feature. For example, the parameter $\lambda$  determines the shape of the potential dip. To some extent, $\lambda$ is manually tuned to ensure that the fraction $f_1(\phi)/(1+f_1 (\phi))$ is appropriate for achieving ultra slow-roll inflation. In our work we set $\lambda = 5\times10^{6}  M_P^{-4}$. For excessively large values of $\lambda$ the inflaton becomes confined within the dip, and inflation turns to be standard slow-roll for too small values. The parameter $v_1$ gives the position of the dip, thus controls when the ultra slow-roll stage happens and the peak scale of the power spectrum. We choose the values of $v_1$ in Tab. \ref{table1} so that the peak of the power spectrum lies in the range $10^{12} Mpc^{-1}\sim10^{14} Mpc^{-1}$. Therefore, the formed primordial black holes have asteroid mass. The parameter $v_2$ tells the depth of the dip. It is related to the slow roll parameter $\eta$ and the duration of the ultra slow-roll stage. In this work the parameter $v_2$ is fine tuned and it sensitively affects the amplitude of the power spectrum.

The field equation of the inflaton is
\be
\ddot{\phi}+3H\dot{\phi}+V_{,\phi}=0,
\ee
where dot represents time derivative and $H=\dot{a}/a$ is the Hubble parameter, with $a$ the scale factor.
One may also write this equation with the e-folding number,
\be
\phi_{,NN}+3\phi_{,N}-\frac{1}{2}\phi_{,N}^{3}
+\left(3-\frac{1}{2}\phi_{,N}^{2}\right)\frac{V_{,\phi}}{V}=0,
\label{beq evolution}
\ee
where the subscripts $(,\phi)$ and $(,N)$ represent the derivative with respect to the inflaton field $\phi$ and the e-folding number $N$, respectively. Note that the e-folding number $N$ and the time coordinate $t$ are related by $Hdt=dN$.
Defining the slow-roll parameters,
\be
\epsilon\equiv -\frac{\dot{H}}{H^2}=\frac{1}{2}\phi_{,N}^2,\quad
\eta\equiv\frac{\dot{\epsilon}}{H\epsilon}=2\frac{\phi_{,NN}}{\phi_{,N}},
\ee
the evolution equation (\ref{beq evolution}) then becomes
\be
\eta=-2(3-\epsilon)\left(1\pm\sqrt{\frac{\epsilon_V}{\epsilon}}\right),
\label{eta_expression}
\ee
\begin{figure}[htb]
  \centering
  \includegraphics[width=6.5cm]{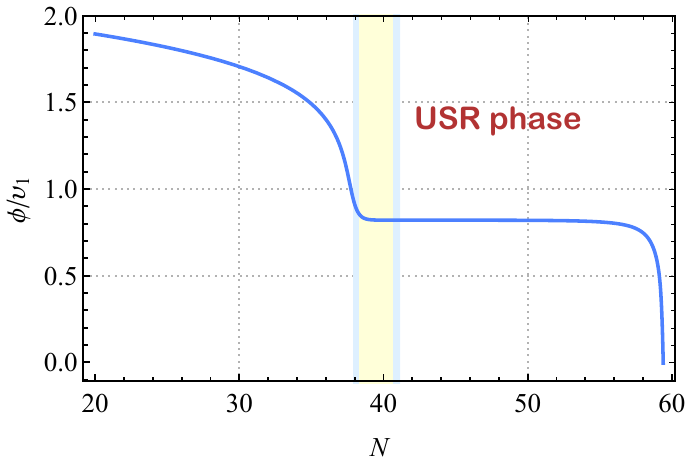}
  \includegraphics[width=6.7cm]{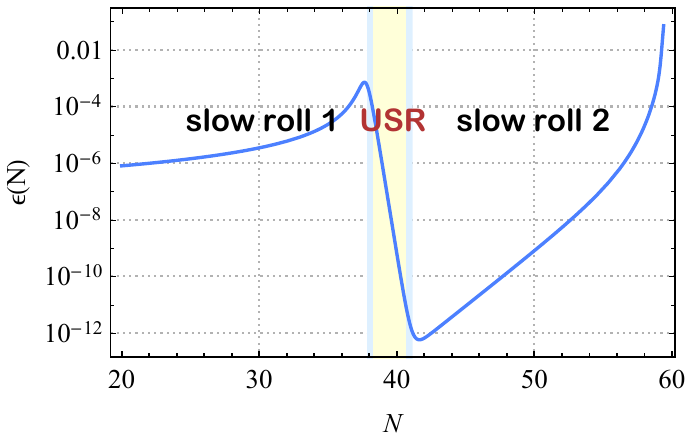}
  \includegraphics[width=6.5cm]{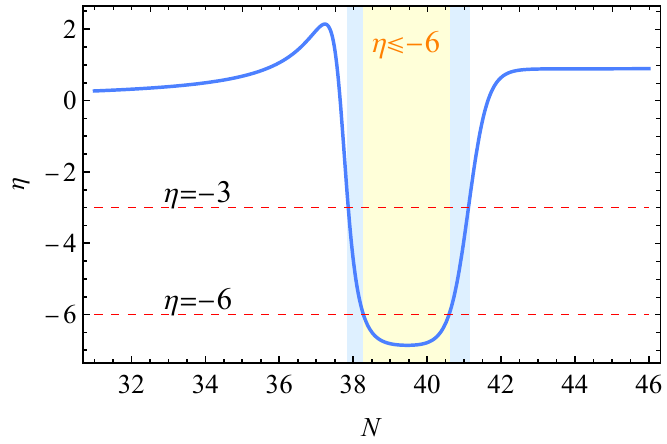}
  \caption{The evolution of the inflaton (upper) and the slow-roll parameters (lower) for parameter set $P_1$ in Table~\ref{table1}. The dashed red lines represent $\eta=-6$ and $-3$, respectively. Their crossing with the numerical $\eta(N)$ define the USR stage (yellow region) and $\eta\leq-3$ stage (blue region).}
  \label{figure1}
\end{figure}
where
\be
\epsilon_V=\frac{1}{2}\left(\frac{V_{,\phi}}{V}\right)^2
\ee
is the potential slow-roll parameter. The ``$\pm$" sign appears because of the nonmonotonicity of the potential in our theory.
For the usual slow-roll inflation, the potential is monotonic (for inflation period) and $V_\phi$ is always positive (or negative) in inflation phase. For USR inflation with a quasi-inflection point (but still monotonic), the potential is extremely flat and the slope $V_{,\phi}\simeq0$. Equation (\ref{eta_expression}) has a ``$-$" sign and straightforwardly gives $\eta\simeq-6$ around the plateau.  In our theory, equation (\ref{eta_expression}) has ``$-$" sign before the inflaton rolls down to the local minimum. After the inflaton passes through the local minimum, equation (\ref{eta_expression}) has a ``$+$" sign before the local maximum. In this period we have $\eta<-6$. As is analyzed in \cite{Gu:2022pbo}, this $\eta<-6$ stage is crucial since the curvature perturbations grow as $\mathcal{R}_k\sim e^{-\eta N/2}$ in the USR stage. In this way, the curvature perturbations and thus the power spectrum are enhanced sufficiently so that the condition of PBH formation is satisfied. The evolution of $\eta$ is shown in Fig. \ref{figure1}.

\section{The curvature perturbations and the power spectrum}\label{sec2}
\subsection{The analytical approach}
The USR stage amplifies the curvature perturbations. To be more specific, let us consider the evolution of the curvature perturbations analytically. The evolution equation of the curvature perturbation modes is
\be
\mathcal{R}_{k,NN}+(3-\epsilon+\eta)\mathcal{R}_{k,N}+\frac{k^2}{a^2H^2}\mathcal{R}_k=0.
\label{curvatureEq}
\ee
Note that $\mathcal{R}$ represents the curvature perturbation and $\mathcal{R}_k$ is its Fourier mode with wave number $k$. It is also convenient to work in the conformal time coordinate and use the Mukhanov-Sasaki variable. In this work however, we deal with the equation (\ref{curvatureEq}) directly. For constant $\epsilon$ and $\eta$, the general solution is
\begin{align}
\mathcal{R}_k(N)=C F^{(1)}_{\nu}\left(\frac{k}{aH}\right)
+D F^{(2)}_{\nu}\left(\frac{k}{aH}\right),
\end{align}
where $F^{(1)}_{\alpha}(x)=\left(x/2\right)^\alpha J_{-\alpha}(x)\Gamma(1-\alpha)$ and $F^{(2)}_{\alpha}(x)=\left(x/2\right)^\alpha J_{\alpha}(x)\Gamma(1+\alpha)$, with $J_{\alpha}(x)$ the Bessel function of the first kind and $\Gamma(x)$ the Gamma function. $C$ and $D$ are coefficients with $k$ dependence, and $\nu=(3-\epsilon+\eta)/2$.

Based on the above observation and the evolution of the slow-roll parameters, we divide the inflation process into four stages, the slow-roll stage
($N<N_1$), the USR stage ($N_1\leq N < N_2$), the
constant-roll stage ($N_2 \leq N < N_3$), and the final slow-roll stage ($N_3\leq N$). We neglect the slow-roll parameter $\epsilon$ for the moment due to its smallness. We will discuss the impact of $\epsilon$ in next subsection. The parameter $\eta$ is approximated as
\be
\eta\simeq\left\{
        \begin{array}{ll}
          0, & \hbox{$0\leq N<N_1$} \\
          \eta_u, & \hbox{$N_1\leq N< N_2$} \\
          \eta_c, & \hbox{$N_2\leq N < N_3$} \\
          0, & \hbox{$N_3\leq N $}
        \end{array}\\
        \right..
        \label{eta_multiphase}
\ee
Note that $N=0$ corresponds to the start of inflation.
\begin{figure}[htb]
  \centering
  \includegraphics[width=6.5cm]{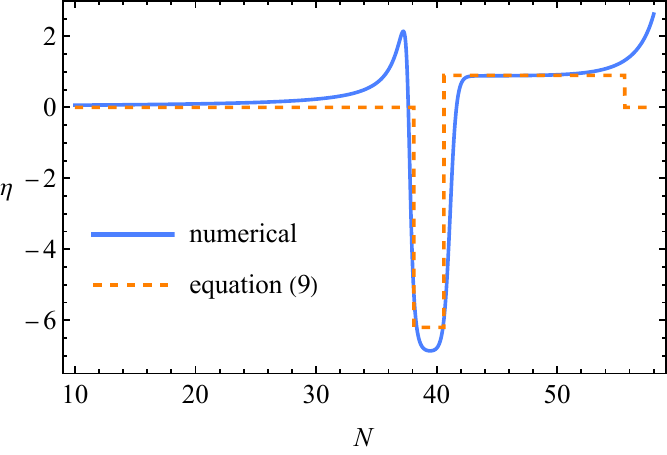}
  \includegraphics[width=7cm]{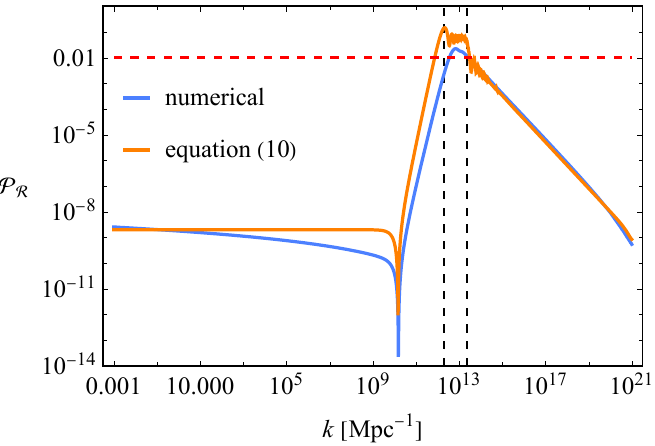}
  \caption{\textit{Upper}: The schematic of the model (\ref{eta_multiphase}) and its comparison to the numerical $\eta$ (for $P_1$ set parameters). The parameters are $N_2-N_1=2.5$, $\eta_u=-6.2$, $N_3-N_2=18$, and $\eta_c=0.9$. The transition $N_1\simeq 38.4$ is set such that the predicted dip of the power spectrum coincide with that of the numerical one. Note that the CMB scales exit the horizon at $N_{\text{CMB}}\simeq 9$. \textit{Lower}: A comparison of the power spectrum obtained by numerical evolution (for parameter set $P_1$) and the analytical one constructed in upper panel. The dashed vertical lines correspond to $k_1=a(N_1)H$ (left) and $k_2=a(N_2)H$ (right). The dashed red line is $\mathcal{P}_{\mathcal{R}}=10^{-2}$, the necessary condition of PBH production.}
  \label{figure_anaPPS}
\end{figure}

From the numerical evolution of the background quantities in Fig.~\ref{figure1}, we see that
the slow-roll parameter $\eta$ starts growing and exceeds unity after the constant-roll stage. This certainly violates the slow-roll condition. However, in (\ref{eta_multiphase}) we assume that for $N\geq N_3$ the inflation process turns back to be slow-roll and $\eta\simeq0$ for simplicity. This approximation is reasonable provided that the constant-roll phase lasts long enough, in which case the scales responsible for PBH production have exited the horizon and being frozen for a long time at $N=N_3$, so that they would not be affected by the final stage ($N\geq N_3$).

Under this setup we can solve the curvature perturbation modes in each stage analytically. The coefficients are obtained by the continuity conditions at transitions. Finally, the power spectrum in the late time limit ($k\ll aH$) is given by
\be
\mathcal{P}_\mathcal{R}(k)=\lim_{k\ll aH}
\frac{k^3}{2\pi^2}\big|\mathcal{R}^{(3)}_{k}\big|^2\simeq
\frac{k^3}{2\pi^2}\left|C_3\right|^2.
\label{powerspectrum}
\ee
The coefficients are computed in the Apendix.
We show the comparison of the numerically calculated  power spectrum \footnote{We use the Bunch-Davis vacuum as the initial conditions.} and the analytical one in Fig.~\ref{figure_anaPPS}. We see that they have some differences. First, there is a discrepancy for the modes that exit the horizon in slow-roll regime ($N\leq N_1$). This is because the analytical approach neglected the evolution of the slow-roll parameter $\epsilon$ in deriving the analytical power spectrum (\ref{powerspectrum}). To understand this, let us consider the expansion of the power spectrum (\ref{powerspectrum}) in $k$,
\begin{align}
\!\!\!\mathcal{P}_\mathcal{R}(k)\!\simeq \!\frac{H^2}{8\pi^2 \epsilon_*}\Bigg[1\!+\!c_2\left(\frac{k}{k_1}\right)^2\!+\!c_4\left(\frac{k}{k_1}\right)^4\!+\!c_6\left(\frac{k}{k_1}\right)^6\!+\!\mathcal{O}(k^8)\Bigg],
\label{expansion_woEp}
\end{align}
\begin{figure}[htb]
  \centering
  \includegraphics[width=7.5cm]{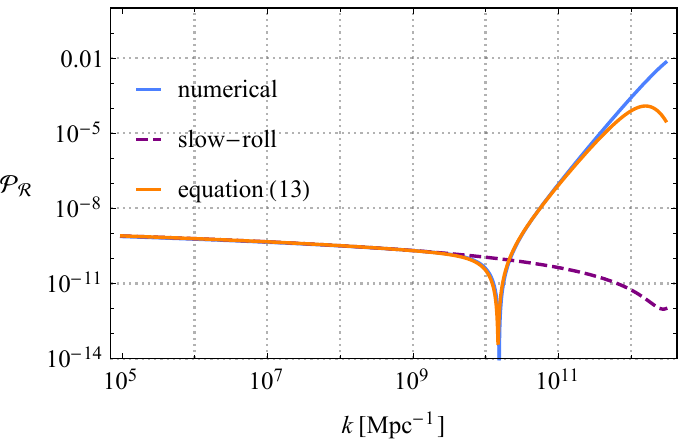}
  \caption{The power spectrum computed by different approaches. The slow-roll power spectrum is computed by $\mathcal{P}_{\mathcal{R}}=\frac{H^2}{8\pi^2\epsilon}$.
  The parameters are the same as those of Fig. \ref{figure_anaPPS}.}
  \label{anaPPS_imp}
\end{figure}
where $\epsilon_*$ is evaluated at the time when the CMB pivot scale $k_*=0.05~\text{Mpc}^{-1}$ exits the horizon. In this equation and the rest of this paper, $k_i=a(N_i)H$. Hence $k_1=a(N_1)H$ represents the scale that exits the horizon at transition $N=N_1$. The coefficients $c_i$ are functions of slow-roll parameters and the duration of the each stage of inflation, hence the effects of super-horizon evolution are included in them. This expansion gives a scale invariant power spectrum for scales $k\ll k_1$, as is shown in Fig.~\ref{figure_anaPPS}. As a rough modification which includes the evolution of $\epsilon$, we replace the constant parameter $\epsilon_*$ by $\epsilon(k)$,
\begin{align}
\!\!\!\mathcal{P}_\mathcal{R}(k)\!\simeq\! \frac{H^2}
{8\pi^2 \epsilon(k)}
\Bigg[1\!+\!c_2\left(\frac{k}{k_1}\right)^2
\!+\!c_4\left(\frac{k}{k_1}\right)^4
\!+\!c_6\left(\frac{k}{k_1}\right)^6\!+\!\mathcal{O}(k^8)\Bigg].
\label{ModPPS}
\end{align}
The slow-roll parameter $\epsilon(k)$ is computed at horizon crossing time $N_e=\log (k/H)$. On the other hand, the series of $k/k_1$ in the above square brackets are derived at the end of inflation so that the super-horizon evolution of the curvature perturbations are included. This implies that the approximation (\ref{ModPPS}) is only valid for $k\ll k_1$.
In this regime the power spectrum can be written as
\begin{equation}
\mathcal{P}_\mathcal{R}(k)\simeq
          \frac{k^3}{2\pi^2}\frac{\epsilon_*}{\epsilon(k)}\left|C_3\right|^2.
\label{imp_PPS}
\end{equation}
This expression is not applicable for the scales $k\gtrsim k_1$.
We show the comparison of (\ref{imp_PPS}) and the numerical one in Fig.~\ref{anaPPS_imp}. We see that this modification gives a more accurate result for the modes $k\ll k_1$. Now the remaining problem is, to what scales the approximation (\ref{imp_PPS}) is applicable. Unfortunately, the exact bound is not known since this expression is not derived rigorously. We will investigate this problem in our future work.

The second difference is the oscillations around the peak, which is absent in the numerical power spectrum. The oscillations are fundamentally linked to the specifics of the transition between the slow roll phase and the ultra-slow-roll (USR) phase. Reference [133] parametrized this transition using an analytical model, comparing smooth and instantaneous transitions. It was found that oscillations vanish if the transition lasts more than approximately one e-fold. In essence, the oscillations are a result of instantaneous (sharp) transitions, such as those described by the analytical approximation (9) in our work, as well as the step model discussed in references [89] and [91].

The third difference is that, the analytical power spectrum has a slightly steeper growth rate before the peak. This is also related to the ignorance of the evolution of $\epsilon$. The expansion (\ref{expansion_woEp}) predicts a growth rate of $\sim k^4$ due to the constant $\epsilon_*$ \cite{Byrnes:2018txb}. This conclusion would be changed if the evolution of $\epsilon$ is considered. As is shown in Fig. \ref{figure1}, the parameter $\epsilon(N)$ grows with $N$ before the transition at $N_1$, hence $\epsilon(k)$ grows with $k$ ($k<k_1$). Therefore the $k$-dependence of the parameter $\epsilon(k)$ slightly reduces the $k^4$ growth of the power spectrum. This is consistent with the numerical results.

At last, the analytical power spectrum predicts a larger amplitude of peak, indicating a deviation from the realistic model. We will discuss this issue in next subsection.

\subsection{The enhancement}
We are interested in how many orders of magnitude the power spectrum could be enhanced, since this is responsible for the PBH production.
For the toy model (\ref{eta_multiphase}), the magnitude of enhancement of the power spectrum is \cite{Inomata:2021uqj,Dalianis:2021iig,Inomata:2021tpx}
\begin{align}
\Delta_{\mathcal{P_R}}\!=\!\log_{10}\frac{\mathcal{P_R}(k_{\text{peak}})}{\mathcal{P_R}(k_*)}
\!\simeq\! \log_{10}\frac{\epsilon(N_1)}{\epsilon(N_2)}
\!=\! -0.434\int_{N_1}^{N_2}\eta dN,
\label{magnitude}
\end{align}
where the factor $0.434\simeq\log_{10}{e}$. However, this estimation would be problematic for realistic models since there are uncertainties in determining the bounds of the integral (\ref{magnitude}). This is because the transitions between slow-roll and ultra slow-roll are non-instantaneous. One may intuitively replace the integral bounds $N_1$ and $N_2$ by $N_{-6_-}$ and $N_{-6_+}$,
\begin{figure}[htb]
  \centering
  \includegraphics[width=7.5cm]{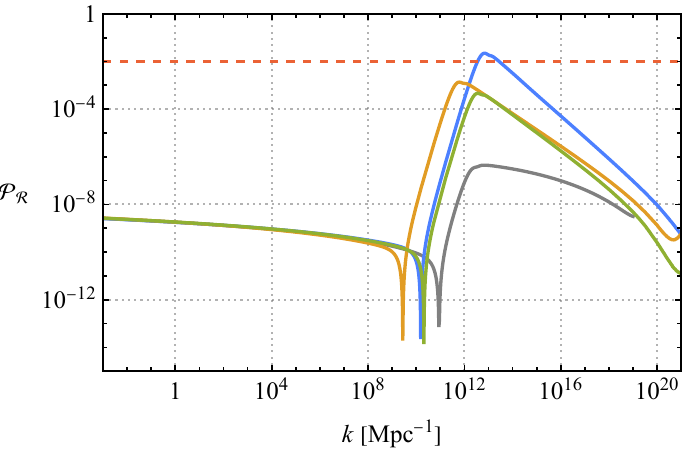}
  \caption{The power spectrum for parameter choices $P_1$ (blue), $P_2$ (orange), $P_3$ (green), and $P_4$ (gray) in Tab.~\ref{Enhancecom}. The dashed red line ($\mathcal{P_{R}}=10^{-2}$) represents the threshold to have considerable amount of PBHs.}
  \label{PPS_com}
\end{figure}
\be
\Delta_{-6}\simeq -0.434\int_{N_{-6_-}}^{N_{-6_+}}\eta dN,
\label{magnitude1}
\ee
where  $N_{-6_{\pm}}$  are the e-folding numbers when $\eta=-6$. This estimation predicts an enhancement close to the exact value for short transitions ($\Delta N\lesssim 1$). Nevertheless, it underestimates the enhancement since the curvature perturbations grow even if $\eta>-6$. Mathematically, the equation (\ref{curvatureEq}) has exponentially growing solution as long as $3-\epsilon+\eta<0$, irrespective of whether the mode is sub-horizon or super-horizon. Neglecting the parameter $\epsilon$, this corresponds to $\eta<-3$. Hence, a conservative estimate for the upper bound of $\Delta_{\mathcal{P_R}}$ is
\be
\Delta_{\text{max}}\simeq -0.434\int_{N_{-3_-}}^{N_{-3_+}}\eta dN,
\label{magnitude2}
\ee
where $N_{-3_{\pm}}$ corresponds to $\eta=-3$. We compare the enhancement predicted by equation (\ref{magnitude1}), (\ref{magnitude2}), and the numerical results in Tab.~\ref{Enhancecom} and Fig. \ref{PPS_com}. We see that the magnitude of enhancement is between the estimate (\ref{magnitude1}) and (\ref{magnitude2}). As expected, the equation (\ref{magnitude1}) gives a close but lower estimate of enhancement. The equation (\ref{magnitude2}) overestimates the enhancement because of the ignorance of the details of each stage. As was pointed out in \cite{Byrnes:2018txb,Cole:2022xqc}, the stages before and after USR and the details of the transitions between different stages have significant effects on the amplitude of the power spectrum. Therefore, an exact estimate for the amplitude of the peak should contain the information of the whole inflation process.

\begin{table}
\centering
\begin{tabular}{c| c c c c}
  \hline\hline
       Sets& ~~$\mathcal{P}_{\mathcal{R}}(k_{\text{peak}})$~~&  ~~$\Delta_{\text{num}}$~~ &~~$\Delta_{-6}$~~& ~~$\Delta_{\text{max}}$~~ \\
  \hline
  ~$P_1$~ & $2.22\times 10^{-2}$ & 7.0& 6.7  & 8.6 \\
  ~$P_2$~ & $1.36\times 10^{-3}$ & 5.8& 5.4  & 7.7 \\
  ~$P_3$~ & $4.60\times 10^{-4}$ & 5.3& 5.1  & 7.2 \\
  ~$P_4$~ & $4.45\times 10^{-7}$ & 2.3& ---  & 4.7 \\
  \hline\hline
\end{tabular}
\caption{The magnitude of enhancement and the estimated lower and upper bounds. Note that the parameter $\eta$ is always larger than $-6$ for the set $P_4$.}
\label{Enhancecom}
\end{table}

\section{PBH formation}\label{sec3}

\begin{figure}[htb]
  \centering
  \includegraphics[width=7.5cm]{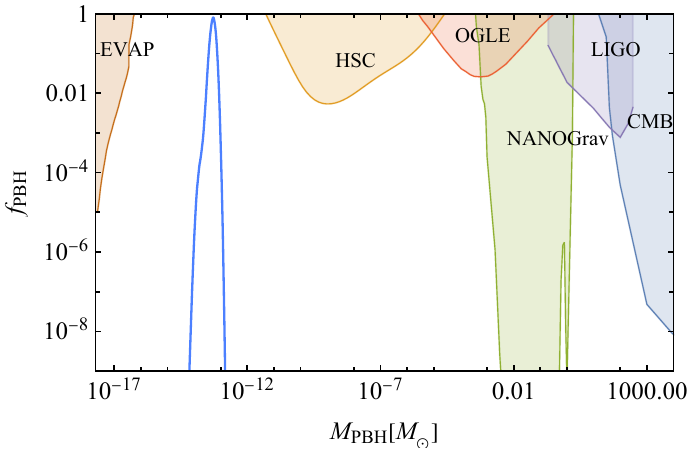}
  \caption{The fraction of PBHs as dark matter for parameter set $P_1$ in Tab.~\ref{Enhancecom}.  The shaded region are the constraints from evaporation of PBHs (EVAP) \cite{Carr:2009jm,Clark:2016nst,Clark:2018ghm,Boudaud:2018hqb,
  Laha:2019ssq,DeRocco:2019fjq,Laha:2020ivk}, Sabaru-HSC (HSC) \cite{Croon:2020ouk}, Optical Gravitational Lensing Experiment (OGLE) \cite{Niikura:2019kqi}, NANOGrav \cite{Chen:2019xse}, LIGO/Virgo \cite{Kavanagh:2018ggo,LIGOScientific:2019kan}, and cosmic microwave background (CMB) \cite{Serpico:2020ehh}.}
  \label{PBHbound}
\end{figure}
PBHs can be formed when the large fluctuations re-enter the horizon.
The mass fraction of PBHs at formation is \cite{Young:2014ana,Sasaki:2018dmp}
\be
\beta(M)=\frac{\rho_{\text{PBH}}}{\rho_\text{tot}}=\gamma\int
_{\delta_c}^{+\infty}\frac{\mathrm{d}\delta}{\sqrt{2\pi}\sigma_{M}}e^{-\frac{\delta^2}{2\sigma^2_{M_\text{PBH}}}},
\ee
where the density fraction $\delta$ is related to the curvature perturbation by
\be
\delta(t,k)=\frac{2(1+\omega)}{5+3\omega}\left(\frac{k}{aH}\right)\mathcal{R}_{k}.
\ee
In this paper the equation of state $\omega=1/3$ since we assume the PBHs and the GWs are formed in radiation domination. The threshold $\delta_c$ is the critical value of density contrast over which the PBHs could be formed. Its value depends on the details of the collapse and still has uncertainties. In this work we set $\delta_c=0.51$ \cite{Young:2019osy}.  The parameter $\gamma=0.2$ is the fraction of mass collapsed to be PBHs that has $\delta>\delta_c$.
$\sigma_{M_{\text{PBH}}}$ is the standard deviation of $\delta$, defined by
\be
\sigma_{M}^2=\int_{0}^{\infty}\frac{\mathrm{d}k}{k}\frac{16}{81}(kR)^4
W^2(kR)\mathcal{P}_{\mathcal{R}}(k),
\ee
where $W(kR)$ is the window function. Due to the increasing of the power spectrum, $W^2(kR)\mathcal{P}_{\mathcal{R}}$ has its maximum at $k\sim1/R$, hence one has $\sigma_{M_\mathrm{PBH}}\sim 4\sqrt{\mathcal{P}_{\mathcal{R}}}/9$. Furthermore, the mass of the produced PBHs are related to the wave number by
\be
M_\mathrm{PBH}\simeq 30M_{\odot}\left(\frac{\gamma}{0.2}\right)\left(\frac{g_{*,\mathrm{f}}}{10.75}\right)^{-\frac{1}{6}}\left(\frac{k}{2.9\times10^5\text{Mpc}^{-1}}\right)^{-2},
\ee
where $M_\odot$ is the solar mass and $g_{*,\mathrm{f}}$ the relativistic degrees of freedom. The fraction of PBHs as dark matter at present time can be expressed as
\begin{align}
\!\!\!f_{\text{PBH}}(M_{\text{PBH}})\!=\!2.7\!\times\!10^8\left(\frac{\gamma}{0.2}\right)^{\frac{1}{2}}
\!\left(\frac{g_{*,\text{f}}}{10.75}\right)^{-\frac{1}{4}}
\!\left(\frac{M_{\text{PBH}}}{M_\odot}\right)^{-\frac{1}{2}}\!\beta(M).
\end{align}
Now, the fraction  $f_{\text{PBH}}$ can be computed by using the power spectrum $\mathcal{P}_{\mathcal{R}}(k)$.

To produce the PBHs with mass in the range
$10^{-16}M_\odot\lesssim M_\text{PBH}\lesssim10^{-13}M_\odot$, in which the PBHs could constitute all the dark matter, the power spectrum should peak at $10^{12}~\text{Mpc}^{-1}\lesssim k \lesssim 10^{14}~\text{Mpc}^{-1}$. In our model, the parameter set $P_1$ gives $k_{\text{peak}}\simeq 6.7\times10^{12}~ \text{Mpc}^{-1}$, leading to the PBHs with $f_{\text{PBH}}\sim 1$ for $M_\text{PBH}\simeq 5.6\times10^{-14}M_\odot$, as is shown in Fig. \ref{PBHbound}.

\begin{figure}[htb]
  \centering
  \includegraphics[width=7.0cm]{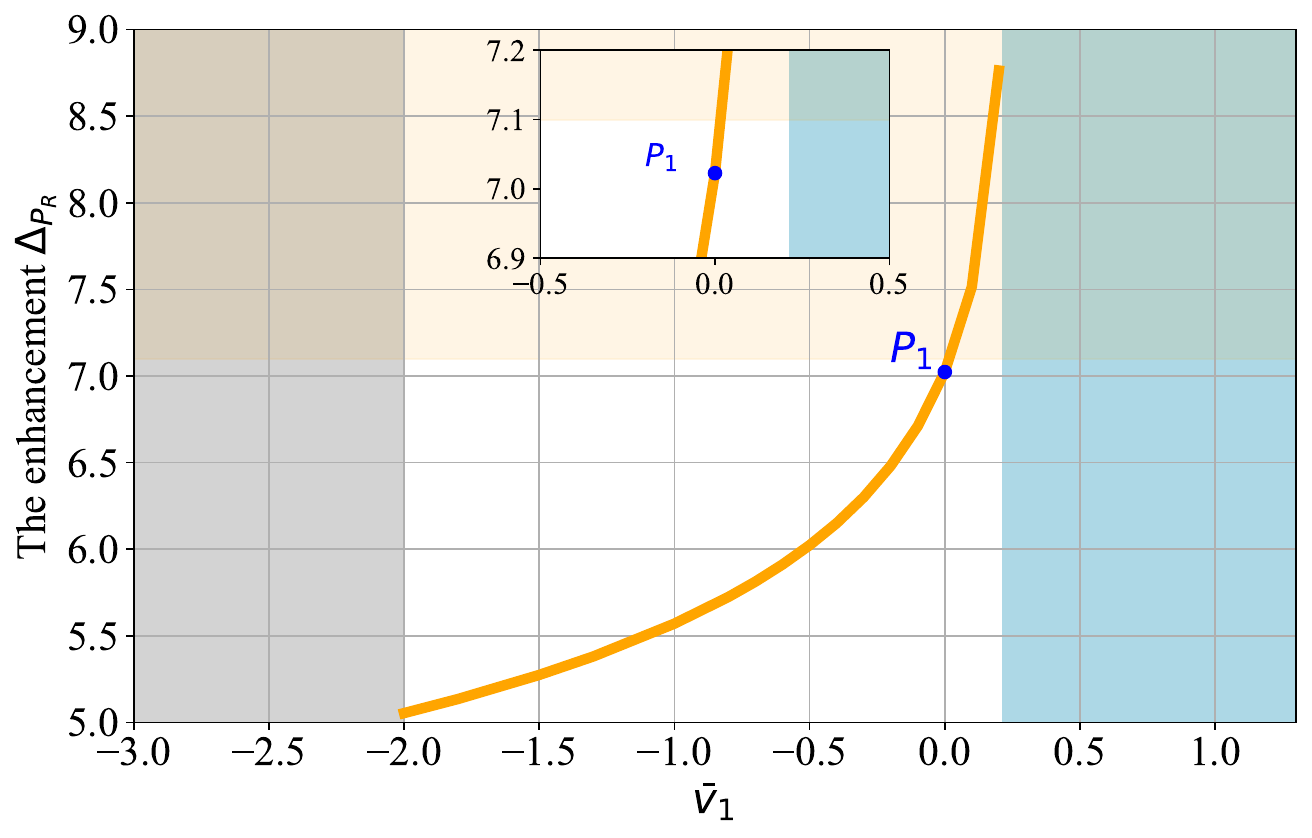}
  \includegraphics[width=7.0cm]{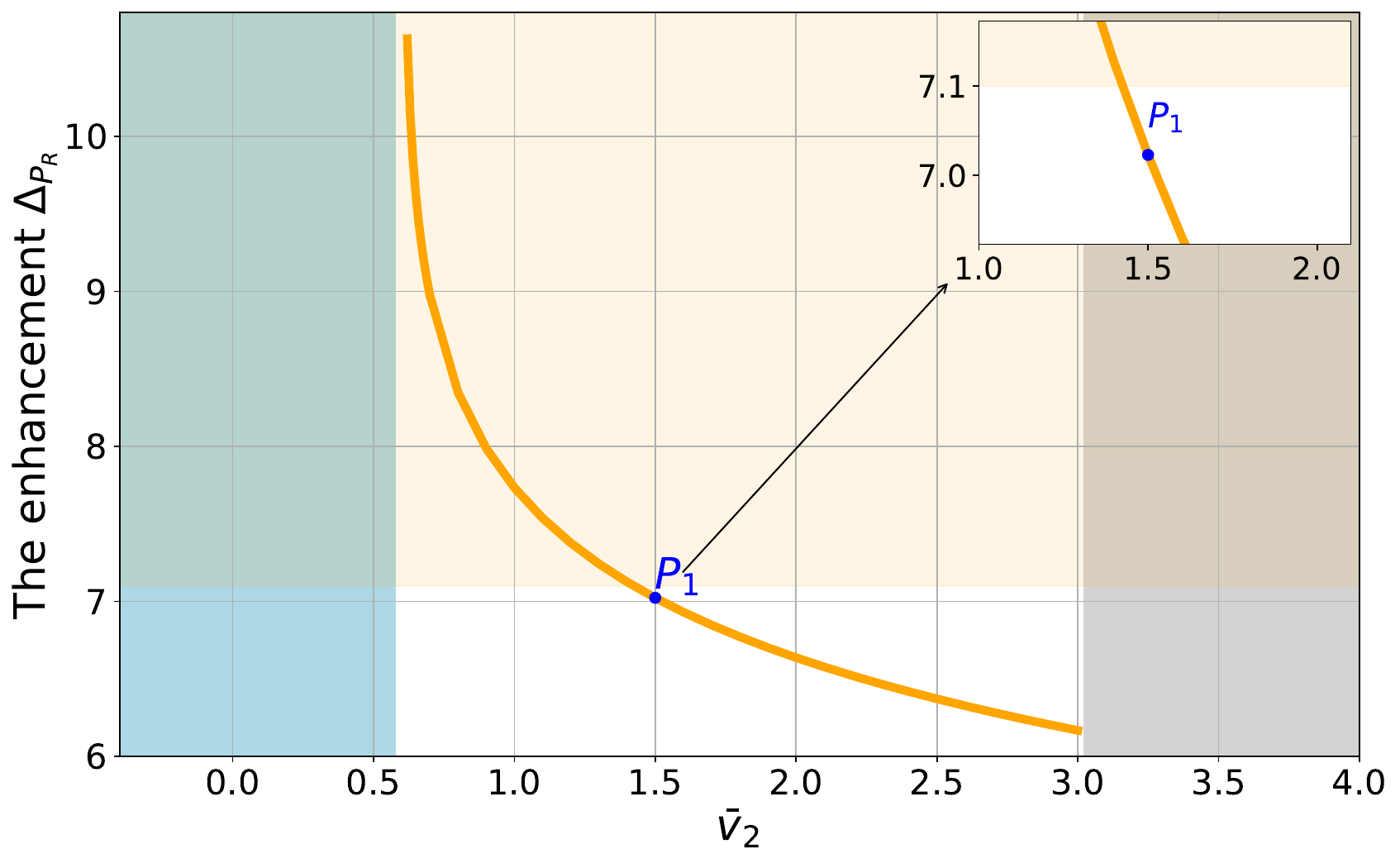}
  \includegraphics[width=7.0cm]{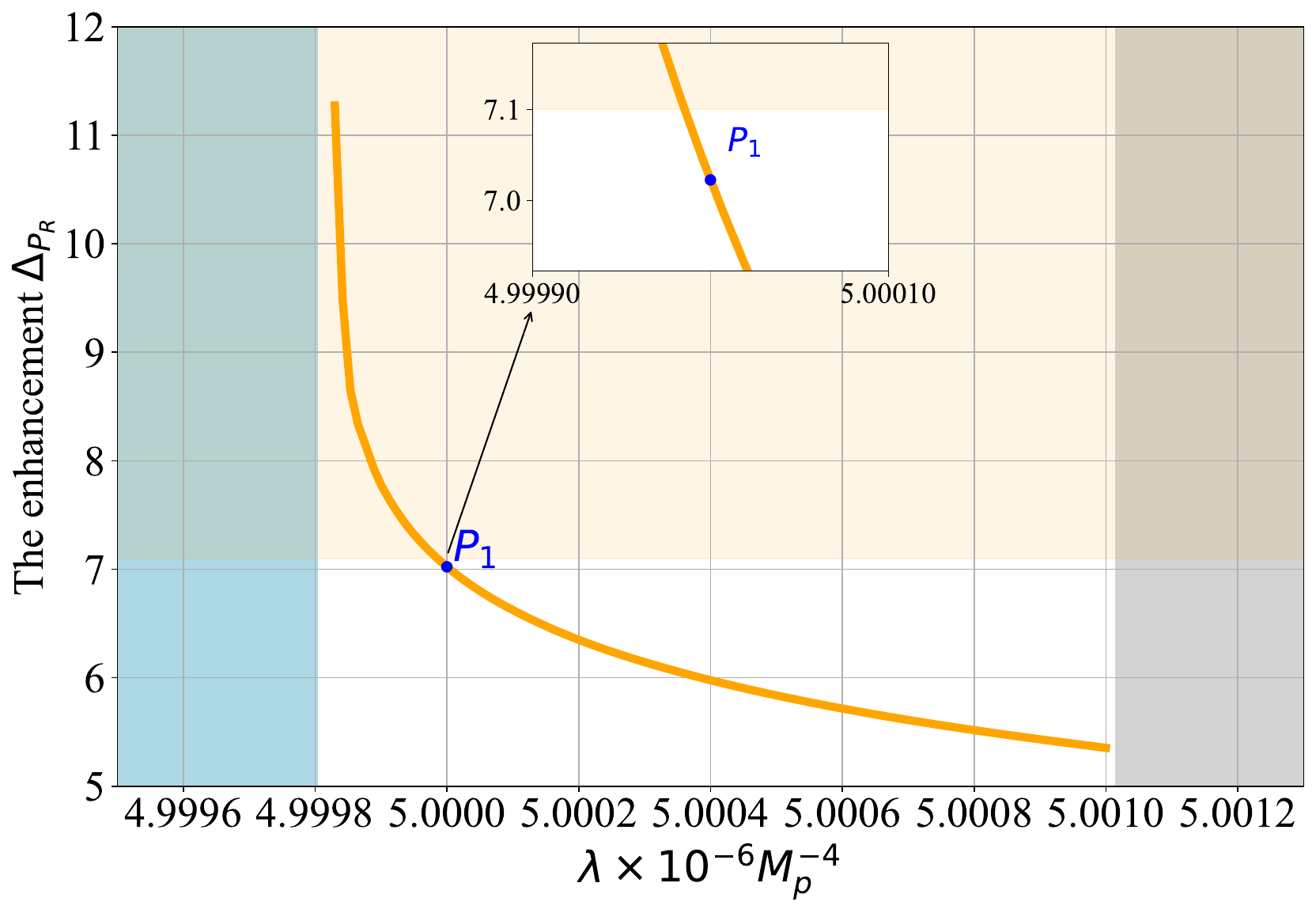}
  \caption{These figures illustrate the relationship between the enhancement of the power spectrum and the parameters $v_1$, $v_2$, and $\lambda$. Note that we use the rescaled axis label $\bar{v}_1=(v_1/M_p-0.094)\times 10^{6}$ and $\bar{v}_2=(v_2/M_p-0.02446)\times 10^7$. All parameter values are identical to those of $P_1$ in Tab. \ref{table1}, except for their respective abscissae. The blue region signifies the parameter space for which the inflation field would be trapped in the potential dip and unable to roll down to the bottom of the potential, while the gray region indicates the parameter space for which the e-folding number after the CMB scales exit is less than 50. The light orange area denotes that the PBHs are over produced due to the large amplitude of power spectrum. The blue-marked point $P_1$ represents exactly the $P_1$ set of parameters in Tab. \ref{table1}.}
  \label{enhance}
\end{figure}

In Fig. \ref{enhance} we show how the parameters $v_1$, $v_2$, and $\lambda$ are related to the enhancement of the power spectrum and the fraction of PBH as dark matter. We choose these parameters because they are directly related to the dip feature of the potential. We study the effects of a specific parameter by fixing the other parameters. Now let us discuss the results. First, we see that there is a fine tuning problem for the parameters. That is, each parameter has a very narrow space that allows for USR. Inappropriate parameters could result in either the inflation field being permanently confined within the potential dip (blue region) or in a total e-folding number of less than 50 (gray region), failing to meet the minimum requirement for inflation. Second, considering the requirements for PBH formation, the range of permissible parameter values will be even narrower. Due to the sensitivity of the exponential dependency, primordial black holes will be overproduced when the enhancement exceeds slightly than 7 (the exact value depends on the details of PBH production, for example the threshold $\delta_c$), whereas when it is slightly less than 7, the abundance of PBHs will be too low to be considerable. These results arise from the exponential dependence of the abundance on the power spectrum. At last, we argue that the parameter space where the peak of the power spectrum exceeds $\sim0.1$ ($\Delta\gtrsim 8$) could exhibit notable nonlinear behavior.  Within such regimes, the linear perturbation theory may break down as the perturbations reach significant magnitudes.

\section{The induced gravitational waves}\label{sec4}
The scalar perturbations decouple with the GWs at linear order. However, they couple at nonlinear orders. Hence, the large scalar perturbations are possible to induce GWs at nonlinear orders. This topic has been widely studied in previous literatures \cite{Ananda:2006af,Baumann2007,Saito:2009jt,Kohri:2018awv,Cai:2018dig,Inomata:2018epa,Domenech:2021ztg}. We follow these works and study the induced GWs in our model. Let us consider the metric
\begin{align}
\!\mathrm{d}s^2\!=\!a^2(\tau)\bigg\{\!-\!\left(1+2\Phi\right)\mathrm{d}\tau^2\!+\!
\bigg[(1-2\Psi)\delta_{ij}
+\frac{1}{2}h_{ij}\bigg]\mathrm{d}x^i\mathrm{d}x^j\bigg\},
\end{align}
where $\Phi$ and $\Psi$ are the first order scalar perturbations and $h_{ij}$ is the second order transverse-traceless tensor perturbation, representing the induced GWs. In the absence of anisotropic stress part, one has $\Phi=\Psi$. In Fourier space, the equation of $h_{ij}$ is given by
\be
h_{\mathbf{k}}''+2\mathcal{H}h_{\mathbf{k}}'+k^2h_{\mathbf{k}}=4S_\mathbf{k},
\ee
with $\mathcal{H}=a'/a$.
Here prime denotes the derivative with respect to the conformal time $\tau$, and $h_\mathbf{k}$ is the Fourier mode of $h_{ij}$ defined by
\be
h_{ij}(\tau,\mathbf{x})=\int\frac{\mathrm{d}^3k}{(2\pi)^{3/2}}
e^{i\mathbf{k}\cdot\mathbf{x}}
[h_{\mathbf{\mathbf{k}}}(\tau)e_{ij}(\mathbf{k})
+\bar{h}_{\mathbf{k}}(\tau)\bar{e}_{ij}(\mathbf{k})],
\ee
where $e_{ij}(\mathbf{k})$ and $\bar{e}_{ij}(\mathbf{k})$ are two unit vector bases perpendicular to $\mathbf{k}$ that are orthogonal to each other.
$S_\mathbf{k}$ is the source term of the induced GWs defined by
\begin{align}
S_\mathbf{k}=&\int\frac{\mathrm{d}^3q}{(2\pi)^{3/2}}
e_{ij}(\mathbf{k})q_iq_j\Bigg[2\Phi_\mathbf{q}\Phi_{\mathbf{k}-\mathbf{q}}
+\frac{4}{3(1+\omega)}
\nonumber\\
&\times\left(\frac{\Phi'_\mathbf{q}}{H}+\Phi_\mathbf{q}\right)
\left(\frac{\Phi'_{\mathbf{k}-\mathbf{q}}}{H}+\Phi_{\mathbf{k}-\mathbf{q}}\right)\Bigg].
\label{equation of hk}
\end{align}
\begin{figure}[htb]
  \centering
  \includegraphics[width=8cm]{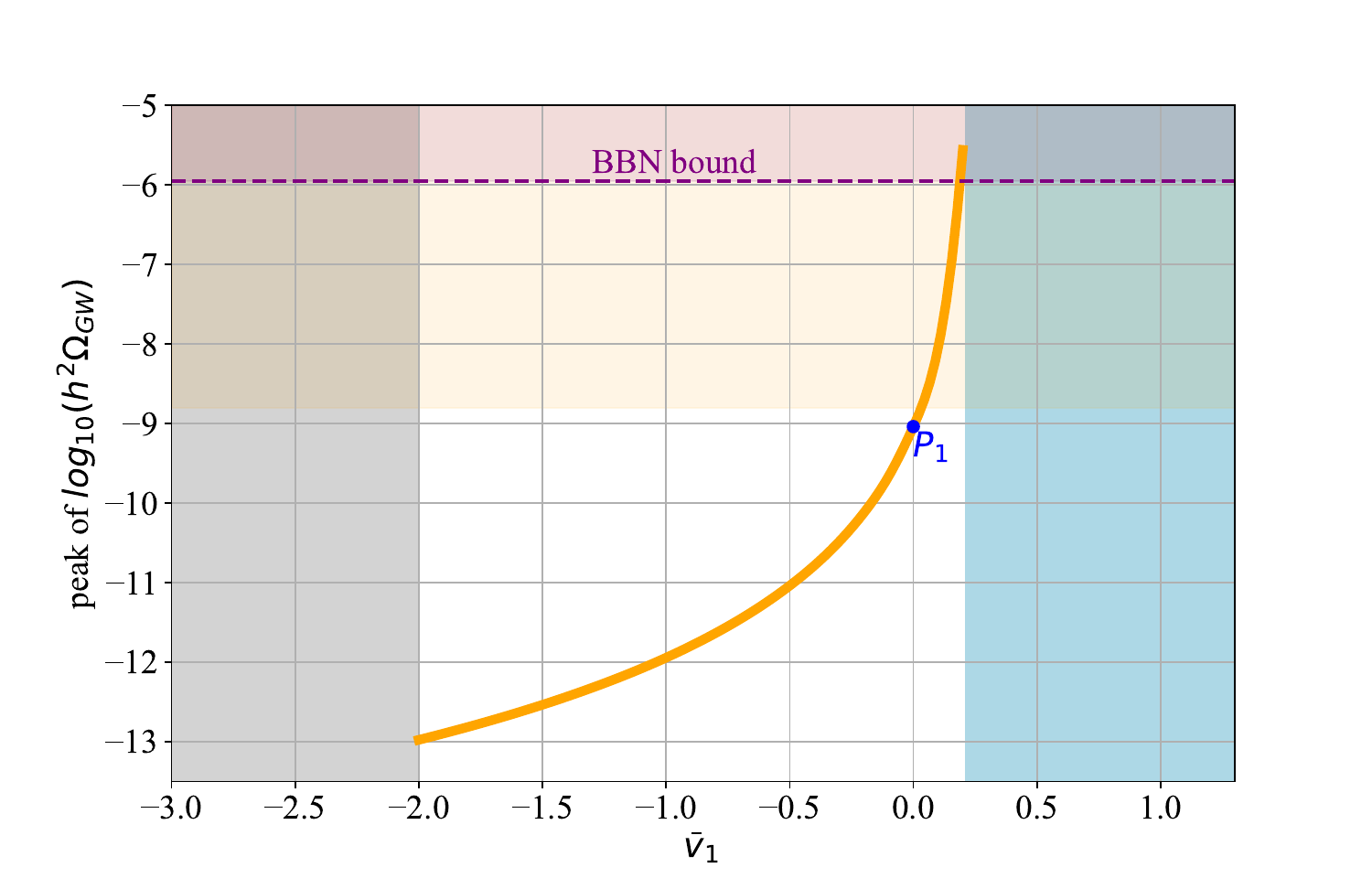}
  \includegraphics[width=8cm]{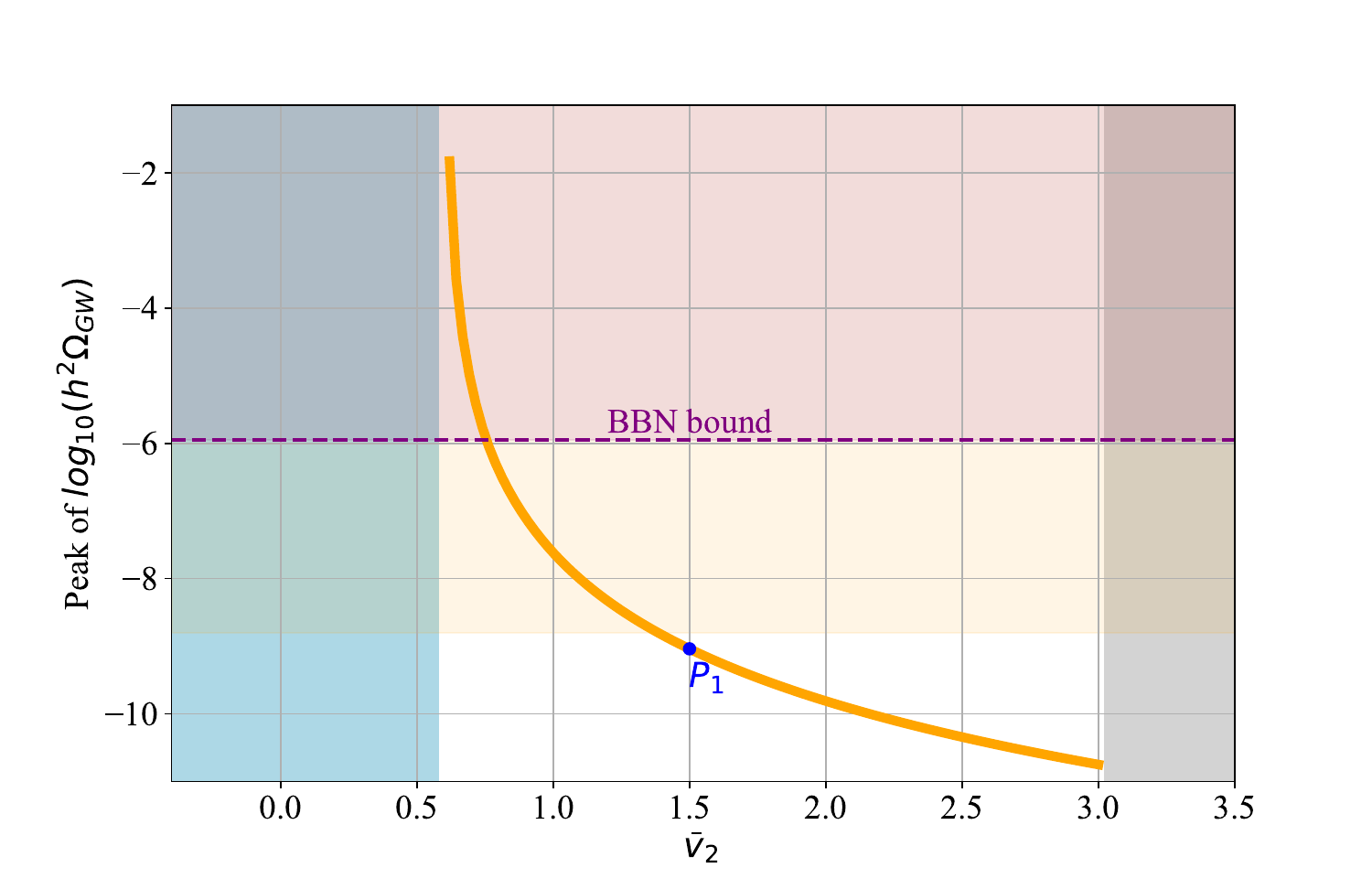}
  \includegraphics[width=8cm]{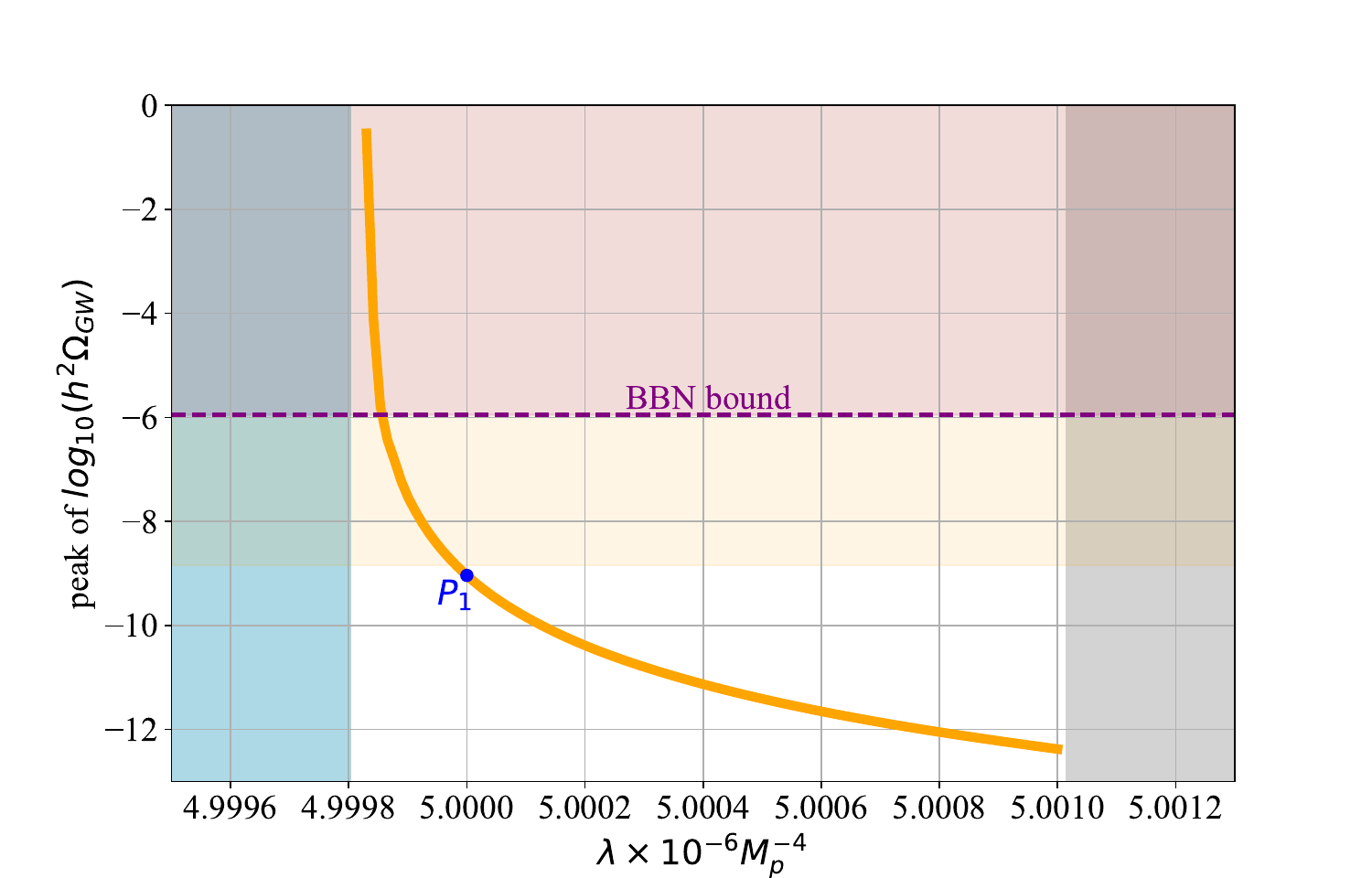}
  \caption{The peak amplitude of the induced GWs versus the parameters $\bar{v}_1$, $\bar{v}_2$, and $\lambda$. The blue, orange, and gray color regions are the same as those in Fig. \ref{enhance}.  The light orange region corresponds to overproduction of PBHs. The purple region represents the constraint from Big Bang Nucleosynthesis (BBN) on the GWs that $h^2 \Omega_{GW}\lesssim 1.12\times 10^{-6}$ \cite{Cyburt:2004yc}. The value of the parameters used are the same as those in Fig. \ref{enhance}. The marked point corresponds to the parameter set $P_1$ in Tab. \ref{table1}.}
  \label{GWsParameter}
\end{figure}
Note that the equation of state parameter $\omega$ is introduced.

The energy density of the GWs in sub-horizon is
\be
\rho_{GW}=\frac{1}{16a^2}\langle\overline{\partial_k h_{ij}\partial^k h^{ij}}\rangle,
\ee
where overline denotes the average over oscillations. The dimensionless power spectrum could be defined by the two-point correlation function,
\be
\langle h_\mathbf{k}^\lambda h_\mathbf{k'}^{\lambda'}\rangle=
\delta_{\lambda\lambda'}\delta^{3}(\mathbf{k}+\mathbf{k}')
\frac{2\pi^3}{k^3}\mathcal{P}_h(\tau,k),
\ee
where the index $\lambda$, $\lambda'$ represent the polarization modes $+$, $\times$.
The energy density of GWs per logarithmic wavelength is
\be
\Omega_{\mathrm{GW}}=\frac{\rho_{\mathrm{GW}}(\tau,k)}{\rho_{\text{tot}}(k)}
=\frac{1}{24}\left(\frac{k}{aH}\right)^2\mathcal{P}_{{h}}(\tau,k).
\ee
The power spectrum of GWs $\mathcal{P}_{{h}}(\tau,k)$ has a lengthy expression. The final result of the energy density at production is \cite{Kohri:2018awv}
\begin{align}
\Omega_{\mathrm{GW,f}}(k)&=\frac{1}{12}\int_0^\infty\!\mathrm{d}\upsilon\!
\int^{1+\upsilon}_{|1-\upsilon|}\!\mathrm{d}u\!
\left(\frac{4\upsilon^2-(1+\upsilon^2-u^2)^2}{4u\upsilon}\right)^2
\nonumber\\
&\times
\mathcal{P}_s(k\upsilon)\mathcal{P}_s(ku)
\left(\frac{3(u^2+\upsilon^2-3)}{4u^3\upsilon^3}\right)^2
\nonumber\\
&\times
\Bigg[\!\left(\!-4u\upsilon+(u^2+\upsilon^2-3)
\log\left|\frac{3-(u+\upsilon)^2}{3-(u-\upsilon)^2}\right|\right)
\nonumber\\
&+
\pi^2(u^2+\upsilon^2-3)^2\Theta(u+\upsilon-\sqrt{3})\Bigg],
\label{GWanaPow}
\end{align}
where $\Theta(x)$ is the step function.
The energy density of GWs at present time is then
\be
\Omega_{\mathrm{GW,0}}=\Omega_{\mathrm{r,0}}\Omega_{\mathrm{GW,f}},
\ee
where $\Omega_{\mathrm{r,0}}$ is the energy density of radiation at present time.
We stress that the formula (\ref{GWanaPow}) is obtained by taking the oscillation average of the trigonometric functions, and the assumption that the GWs are produced in radiation dominated era. If the GWs were produced in matter dominated era, this formula would be modified \cite{Kohri:2018awv}.

\begin{figure}[htb]
  \centering
  \includegraphics[width=8cm]{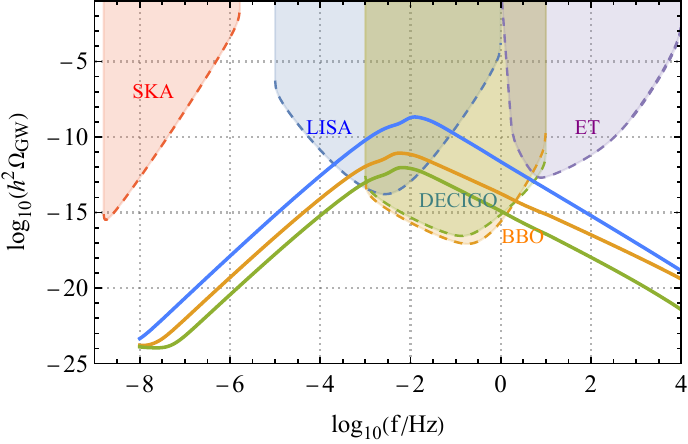}
  \caption{The energy density of the induced GWs for parameter sets $P_1$ (blue), $P_2$ (orange), and $P_3$ (green) against the sensitivity curves of future detectors. The sensitivity curves of the GW detectors are based on the publicly available data \cite{SensCurve}.}
  \label{GWsED}
\end{figure}

Using the numerical power spectrum, the energy density of the induced GWs can be obtained. The Fig. \ref{GWsParameter} shows how the dip feature parameters $v_1$, $v_2$, and $\lambda$ are related to the peak amplitude of the induced GWs. The results of model parameters in Tab. \ref{table1} are shown in Fig. \ref{GWsED}. An interesting feature is that, the parameter sets $P_2$ and $P_3$, whose corresponding PBH abundance is exponentially suppressed since the peak of the power spectrum is not large enough, lead to detectable stochastic GWs by future observatories. This is because the induced GWs are not the product of PBHs, but are of the large perturbations. This confirms the idea that the non-detection of stochastic GWs would strongly constrain the early universe. Recently, the pulsar timing array (PTA) collaborations reported strong evidence for the presence of stochastic GW background \cite{Xu:2023wog,NANOGrav:2023gor,NANOGrav:2023hde,
Reardon:2023gzh,Antoniadis:2023rey}, possibly from supermassive black-hole binaries, but the possibility of other sources like scalar induced GWs are not excluded. If they are induced GWs, they would give a lot of new implications for the primordial curvature perturbations and PBHs \cite{NANOGRAV:2018hou,Ashoorioon:2022raz,Cai:2023dls,Inomata:2023zup,
Depta:2023qst,Franciolini:2023pbf,Liu:2023ymk,Firouzjahi:2023lzg}.

\section{Discussion and conclusions}\label{sec5}
In this work we studied the inflation model proposed in our previous work \cite{Gu:2022pbo}.
The model features a shallow local minimum. Inflation transitions to USR stage when the inflaton rolls over the local minimum. In this stage the perturbations and thus the power spectrum are amplified. For appropriate parameter choices, it is possible to produce PBHs that contribute a large fraction to the dark matter and be consistent with the observations on CMB scales simultaneously. The induced second order GWs are one of the indirect probes of PBHs. The GWs induced by the large perturbations that produce the asteroid mass PBHs are detectable by future experiments like LISA, DECIGO, etc. If these GWs are not observed in future experiments, one would have strong constraints on the primordial scalar power spectrum.

An important issue discussed in this work is the analytical power spectrum. The numerical solution gives the exact power spectrum while the analytical expression would reveal how the power spectrum is controlled by the model parameters. Due to the exponential dependence between the PBH fraction and the power spectrum, a tiny difference of power spectrum would lead to a large deviation of prediction for PBHs.
Therefore, an accurate analytical power spectrum would be important and helpful for model building. We expect to give a complete and more accurate power spectrum in future work.

\section*{Acknowledgements}
This work is supported in part by the National Natural Science Foundation of China (Grants No. 12165013, 11975116, 12005174, 12375049, and 12475062). B.-M. Gu is  supported by Jiangxi Provincial Natural Science Foundation under Grant No. 20224BAB211026. F.-W. Shu is supported by the Key Program of the Natural Science Foundation of Jiangxi Province under Grant No. 20232ACB201008.  K. Yang acknowledges the support of Natural Science Foundation of Chongqing (Grant No.~CSTB2024NSCQ-MSX0358).

\appendix
\section{The coefficients $C_i$ and $D_i$}\label{coeff}

For $0\leq N<N_1$, the slow-roll conditions are satisfied and we have
\beq
\mathcal{R}_k(N)=C F^{(1)}_{\nu}\left(\frac{k}{aH}\right)
+D F^{(2)}_{\nu}\left(\frac{k}{aH}\right),
\eeq
By imposing the Bunch-Davis initial condition, the coefficients $C_0$ and $D_0$ can be determined,
\be
C_0=\frac{i H}{2\sqrt{\epsilon_*k^3}},\quad
D_0=-\frac{4H}{3\sqrt{\epsilon_*k^3}}.
\ee
Hence
\be
\mathcal{R}^{(0)}_k(N)=i\frac{H}{2\sqrt{\epsilon_*k^3}}
e^{i\frac{k}{aH}}\left(1-i\frac{k}{aH}\right).
\ee
In the USR stage ($N_1\leq N< N_2$), the solution is
\beq
\mathcal{R}^{(1)}_k(N)=C_1 F^{(1)}_{\nu_u}\left(\frac{k}{aH}\right)
+D_1 F^{(2)}_{\nu_u}\left(\frac{k}{aH}\right),
\eeq
with $\nu_u\simeq(3+\eta_u)/2$.
The coefficients $C_1$ and $C_2$ are determined by the continuity conditions of $\mathcal{R}_k$ and $\mathcal{R}'_k$ at $N_1$, namely,
\be
\mathcal{R}^{(0)}_k=\mathcal{R}^{(1)}_k\big|_{N=N_1},\quad
\mathcal{R'}^{(0)}_k=\mathcal{R'}^{(1)}_k\big|_{N=N_1}.
\ee
The result is
\begin{align}
C_1=&
-\frac{i H}{2\sqrt{k^3\epsilon_*}}
\frac{\frac{k^2}{k_1^2}F_{\nu_u}^{(2)}\left(\frac{k}{k_1}\right)+
\left(1-i \frac{k}{k_1}\right)F_{\nu_u}'^{(2)}\left(\frac{k}{k_1}\right)}
{W\left[F_{\nu_u}^{(2)}\left(\frac{k}{k_1}\right),
F_{\nu_u}^{(1)}\left(\frac{k}{k_1}\right)\right]}e^{i\frac{k}{k_1}}
\\ \nonumber
=&
\frac{H}{8\sqrt{k^3\epsilon_*}}
\frac{\pi}{\Gamma\left(-\frac{1+\eta_u}{2}\right)}
\sec\left(\frac{\eta_u}{2}\pi\right)e^{i\frac{k}{k_1}}
\bigg\{2i\left[\frac{k^2}{k_1^2}\!+\!(3\!+\!\eta_u)
\left(i\frac{k}{k_1}\!-\!1\right)\right]
\\ \nonumber
&\times{}_0F_1\left(\frac{5+\eta_u}{2},-\frac{k^2}{4k_1^2}\right)+\frac{k^2}{k_1^2}\left[i+\frac{k}{k_1}
{}_0F_1\left(\frac{7+\eta_u}{2},-\frac{k^2}{4k_1^2}\right)\right]\bigg\},
\\
D_1=&
-\frac{i H }{2\sqrt{k^3\epsilon_*}}
\frac{\frac{k^2}{k_1^2}F_{\nu_u}^{(1)}\left(\frac{k}{k_1}\right)+
\left(1-i \frac{k}{k_1}\right)F_{\nu_u}'^{(1)}\left(\frac{k}{k_1}\right)}
{W\left[F_{\nu_u}^{(1)}\left(\frac{k}{k_1}\right),
F_{\nu_u}^{(2)}\left(\frac{k}{k_1}\right)\right]}e^{i\frac{k}{k_1}}
\\ \nonumber
=&\!-\!\frac{H}{2\sqrt{k^3\epsilon_*}}
\frac{\pi}{\Gamma\left(\frac{5+\eta_u}{2}\right)}
\sec\left(\frac{\eta_u}{2}\pi\right)
e^{i\frac{k}{k_1}}\left(\frac{k}{2k_1}\right)^{-\frac{1+\eta_c}{2}}
\!\left[i\frac{k}{k_1}J_{-\frac{3+\eta_u}{2}}\left(\frac{k}{k_1}\right)\right.\\ \nonumber
&\left.+\left(i+\frac{k}{k_1}\right)
J_{-\frac{1+\eta_u}{2}}\left(\frac{k}{k_1}\right)\right],
\end{align}
where $\nu_u\simeq(3+\eta_u)/2$, $W[f,g]=fg'-f'g$ is the Wronskian, $\Gamma(x)$ is the Gamma function, and ${}_0F_1(a,x)$ is the regularized hypergeometric function. After the USR stage, the inflaton undergoes a constant-roll phase ($N_2\leq N\leq N_3$), in which the curvature perturbation is
\begin{align}
\mathcal{R}^{(2)}_k(N)=C_2 F^{(1)}_{\nu_c}\left(\frac{k}{aH}\right)
+D_2 F^{(2)}_{\nu_c}\left(\frac{k}{aH}\right),
\end{align}
with $\nu_c\simeq{3+\eta_c}/2$. The coefficients $C_2$ and $D_2$ can be obtained by repeating the above matching conditions at $N_2$,
\begin{align}
C_2=&
\frac{W\left[C_1 F^{(1)}_{\nu_u}\left(\frac{k}{k_2}\right)+
D_1 F^{(2)}_{\nu_u}\left(\frac{k}{k_2}\right), F^{(2)}_{\nu_c}\left(\frac{k}{k_2}\right) \right]}{W\left[F_{\nu_c}^{(1)}\left(\frac{k}{k_2}\right),
F^{(2)}_{\nu_c}\left(\frac{k}{k_2}\right)\right]}
\\ \nonumber
=&
\frac{\pi}{4\Gamma\left(-\frac{1+\eta_c}{2}\right)}
\sec\left(\frac{\eta_c}{2}\pi\right)
\left(\frac{k}{2k_2}\right)^{\frac{1}{2}(\eta_u-\eta_c)}\!\!
\bigg\{D_1\Gamma\left(\frac{5\!+\!\eta_u}{2}\right)
J_{\frac{5+\eta_c}{2}}\left(\frac{k}{k_2}\right)
\\ \nonumber
&\times J_{\frac{3+\eta_u}{2}}\left(\frac{k}{k_2}\right)
-\frac{k}{k_2}J_{\frac{1+\eta_c}{2}}\left(\frac{k}{k_2}\right)
\left[2C_1\Gamma\left(-\frac{1+\eta_u}{2}\right)
J_{-\frac{3+\eta_u}{2}}\left(\frac{k}{k_2}\right)\right.
\\ \nonumber
&\left.+D_1\Gamma\left(\frac{5+\eta_u}{2}\right)
J_{\frac{3+\eta_u}{2}}\left(\frac{k}{k_2}\right)\right]
+J_{\frac{3+\eta_c}{2}}\left(\frac{k}{k_2}\right)
\bigg[D_1\Gamma\left(\frac{5+\eta_u}{2}\right)
\\ \nonumber
&\left[(3+2\eta_u-\eta_c)J_{\frac{3+\eta_u}{2}}\left(\frac{k}{k_2}\right)
-2\frac{k}{k_2}J_{\frac{5+\eta_u}{2}}\left(\frac{k}{k_2}\right)\right]
\\ \nonumber
&
-2C_1\Gamma\left(-\frac{1+\eta_u}{2}\right)
\frac{k}{k_2}J_{-\frac{1+\eta_u}{2}}\left(\frac{k}{k_2}\right)
\bigg]
\bigg\},
\\
D_2=&
\frac{W\left[C_1 F^{(1)}_{\nu_u}\left(\frac{k}{k_2}\right)+
D_1 F^{(2)}_{\nu_u}\left(\frac{k}{k_2}\right), F^{(1)}_{\nu_c}\left(\frac{k}{k_2}\right) \right]}{W\left[F_{\nu_c}^{(2)}\left(\frac{k}{k_2}\right),
F^{(1)}_{\nu_c}\left(\frac{k}{k_2}\right)\right]}
\\ \nonumber
=&
\frac{\pi}{4}\sec\left(\frac{\eta_c}{2}\pi\right)
\left(\frac{k}{2k_2}\right)^{\frac{\eta_u-\eta_c}{2}}
\Bigg\{D_1\frac{k}{k_2}
J_{-\frac{5+\eta_c}{2}}\left(\frac{k}{k_2}\right)
J_{\frac{3+\eta_u}{2}}\left(\frac{k}{k_2}\right)
\\ \nonumber
&
\!-\!\frac{k}{k_2}J_{-\frac{1+\eta_c}{2}}\left(\frac{k}{k_2}\right)
\left[2C_1\frac{\Gamma\left(-\frac{1+\eta_u}{2}\right)}
{\Gamma\left(\frac{5+\eta_c}{2}\right)}
J_{-\frac{3+\eta_u}{2}}\left(\frac{k}{k_2}\right)
\!+\!D_1J_{\frac{3+\eta_u}{2}}\left(\frac{k}{k_2}\right)\right]
\\ \nonumber
&
+J_{-\frac{3+\eta_c}{2}}\left(\frac{k}{k_2}\right)
\bigg[2C_1\frac{\Gamma\left(-\frac{1+\eta_u}{2}\right)}
{\Gamma\left(\frac{5+\eta_c}{2}\right)}
\frac{k}{k_2}
J_{-\frac{1+\eta_u}{2}}\left(\frac{k}{k_2}\right)
\\ \nonumber
&+D_1(\eta_c-2\eta_u-3)J_{\frac{3+\eta_u}{2}}\left(\frac{k}{k_2}\right)
+2D_1\frac{k}{k_2}J_{\frac{5+\eta_u}{2}}\left(\frac{k}{k_2}\right)
\bigg]
\Bigg\}.
\end{align}
In the final slow-roll stage we have
\begin{align}
\mathcal{R}^{(3)}_k(N)=C_3 F^{(1)}_{3/2}\left(\frac{k}{aH}\right)
+D_3 F^{(2)}_{3/2}\left(\frac{k}{aH}\right).
\end{align}
Note that we assumed that $\eta=0$ and $\nu=3/2$ in this stage.
The coefficients $C_3$ and $D_3$ are given by the matching at $N_3$,
\begin{align}
C_3=&
\frac{W\left[C_2F^{(1)}_{\nu_c}\left(\frac{k}{k_3}\right)
+D_2 F^{(2)}_{\nu_c}\left(\frac{k}{k_3}\right),
F^{(2)}_{3/2}\left(\frac{k}{k_3}\right)\right]}
{W\left[F^{(1)}_{3/2}\left(\frac{k}{k_3}\right),
F^{(2)}_{3/2}\left(\frac{k}{k_3}\right)\right]}
\\ \nonumber
=&
\frac{1}{2}\left(\frac{k}{2k_3}\right)^{\frac{1+\eta_c}{2}}
\Bigg\{
C_2 \Gamma\left(-\frac{1+\eta_c}{2}\right)
J_{-\frac{3+\eta_c}{2}}\left(\frac{k}{k_3}\right)\sin \frac{k}{k_3}
\\ \nonumber
&\!+\!\bigg[C_2\Gamma\left(-\frac{1+\eta_c}{2}\right)
J_{-\frac{1+\eta_c}{2}}\left(\frac{k}{k_3}\right)
+D_2 \Gamma\left(\frac{5+\eta_c}{2}\right)
J_{\frac{5+\eta_c}{2}}\left(\frac{k}{k_3}\right)\bigg]
\\ \nonumber
&\times \left(\frac{k_3}{k}\sin \frac{k}{k_3}-\cos \frac{k}{k_3}\right)
+D_2 \Gamma\left(\frac{5+\eta_c}{2}\right)
J_{\frac{3+\eta_c}{2}}\left(\frac{k}{k_3}\right)
\\ \nonumber
&\times \left[(3+\eta_c)\frac{k_3}{k}\cos \frac{k}{k_3}
+\left(1-(3+\eta_c)\frac{k_3^2}{k^2}\right)\sin \frac{k}{k_3}
\right]
\Bigg\},
\\
D_3=&
\frac{W\left[C_2F^{(1)}_{\nu_c}\left(\frac{k}{k_3}\right)
+D_2 F^{(2)}_{\nu_c}\left(\frac{k}{k_3}\right),
F^{(1)}_{3/2}\left(\frac{k}{k_3}\right)\right]}
{W\left[F^{(2)}_{3/2}\left(\frac{k}{k_3}\right),
F^{(1)}_{3/2}\left(\frac{k}{k_3}\right)\right]}
\\ \nonumber
=&
\frac{4}{3}\left(\frac{k}{2k_3}\right)^{\frac{1+\eta_c}{2}}
\!\!\!\Bigg\{
d_2 \Gamma\left(\frac{5\!+\!\eta_c}{2}\right)
J_{\frac{3+\eta_c}{2}}\left(\frac{k}{k_3}\right)
\left[\left((3\!+\!\eta_c)\frac{k_3^2}{k^2}\!-\!1\right)\cos \frac{k}{k_3} \right.\\ \nonumber
&
\left.+(3+\eta_c)\frac{k_3}{k}\sin \frac{k}{k_3}
\right]
-C_2 \Gamma\left(-\frac{1+\eta_c}{2}\right)
J_{-\frac{3+\eta_c}{2}}\left(\frac{k}{k_3}\right)\cos \frac{k}{k_3}
\\ \nonumber
&-\bigg[C_2 \Gamma\left(-\frac{1+\eta_c}{2}\right)
J_{-\frac{1+\eta_c}{2}}\left(\frac{k}{k_3}\right)
\!+\!D_2 \Gamma\left(\frac{5+\eta_c}{2}\right)
J_{\frac{5+\eta_c}{2}}\left(\frac{k}{k_3}\right)\bigg]
\\ \nonumber
&
\times \left(\frac{k_3}{k}\cos \frac{k}{k_3}+\sin \frac{k}{k_3}\right)
\Bigg\}.
\end{align}
Using the asymptotic property of the Bessel functions, we have
\begin{align}
\lim_{k\ll aH} F^{(1)}_{3/2}\left(\frac{k}{aH}\right)=1,
\quad \lim_{k\ll aH} F^{(2)}_{3/2}\left(\frac{k}{aH}\right)=0.
\end{align}
Thus, the power spectrum in the late time limit is given by
\begin{align}
\mathcal{P}_\mathcal{R}(k)=\frac{k^3}{2\pi^2}
\big|\mathcal{R}^{(3)}_k \big|_{k\ll aH}^2\simeq
\frac{k^3}{2\pi^2}\big|C_3\big|^2.
\end{align}

\end{document}